\documentclass[final,1p,number,sort&compress]{elsarticle}

\usepackage{amssymb}
\usepackage{natbib}
\usepackage{color}
\usepackage{graphicx}
\usepackage{graphics}
\usepackage{dcolumn}
\usepackage{bm}
\usepackage{amsmath}    
\usepackage{verbatim}   
\usepackage{subfigure}  
\usepackage{hyperref}   

\newtheorem{thm}{Theorem}[section]
\newtheorem{cor}[thm]{Corollary}

\newtheorem{lem}[thm]{Lemma}
\newtheorem{prop}[thm]{Proposition}
\newtheorem{defn}[thm]{Definition}
\newtheorem{example}[thm]{Example}
\newtheorem{oss}[thm]{Remark}
\newtheorem{fig}[thm]{Figure}

\def\rl{\mathop{\operatorname{R}_\lambda}}
\def\re{\mathop{\operatorname{{Re}}}}

\newcommand{\bthe}{\begin{thm}}
\newcommand{\ethe}{\end{thm}}

\newcommand{\ble}{\begin{lem}}
\newcommand{\ele}{\end{lem}}

\newcommand{\bde}{\begin{defn}}
\newcommand{\ede}{\end{defn}}

\newcommand{\bco}{\begin{cor}}
\newcommand{\eco}{\end{cor}}

\newcommand{\bpr}{\begin{prop}}
\newcommand{\epr}{\end{prop}}

\newcommand{\bproof}{\begin{proof}}
\newcommand{\eproof}{\end{proof}}

\newcommand{\bexam}{\begin{example}\rm}
\newcommand{\eexam}{\halmos\end{example}}

\newcommand{\brem}{\begin{oss}\rm}
\newcommand{\erem}{\halmos\end{oss}}

\newcommand{\bfi}{\begin{fig}}
\newcommand{\efi}{\end{fig}}

\newcommand{\beao}{\begin{eqnarray*}}
\newcommand{\eeao}{\end{eqnarray*}\noindent}

\newcommand{\beam}{\begin{eqnarray}}
\newcommand{\eeam}{\end{eqnarray}\noindent}

\newcommand{\barr}{\begin{array}}
\newcommand{\earr}{\end{array}}

\newcommand{\beq}{\begin{equation}}
\newcommand{\eeq}{\end{equation}}

\def\bbr{{\Bbb R}}

\newcommand{\CMA}{{\rm CMA}}

\def\halmos{\hfill $\Box$  \medskip }

\begin{document}

\begin{frontmatter}



\title{Turbulence modeling by time-series methods}


\author[rvt]{Vincenzo Ferrazzano}
\ead{ferrazzano@ma.tum.de}
\author[rvt,focal]{Claudia Kl\"uppelberg}
\ead{cklu@ma.tum.de}
\address[rvt]{Center for Mathematical Sciences, Technische Universit\"at M\"unchen, D-85747 Garching, Germany}
\address[focal]{Institute for Advanced Study, Technische Universit\"at M\"unchen}
\begin{abstract}
A general model for stationary, time-wise turbulent velocity
is presented and discussed.
This approach, inspired by modeling ideas of {\cite{barndorff-nielsen:372}}, is coherent with the K41 hypothesis of local isotropy, and it allows us to separate second-order statistics from higher order ones.
{The model can be motivated} by Taylor's hypothesis and a relation between time and spatial spectra.
Second order statistics are used to separate the deterministic kernel function and the weakly stationary driving noise.
A non-parametric estimation method for the turbulence intermittency is suggested.
\begin{description}
\item[Available at]{http://www-m4.ma.tum.de/en/research/preprints-publications/}
\end{description}
\end{abstract}


\end{frontmatter}



\section{Introduction}\label{sec:experimental}

The Wold-Karhunen representation (\cite{Doob:1990df}, p.~588)  states that every non-deterministic, one dimensional, stationary stochastic process $Y=\{Y_{h}\}_{h\in\bbr}$,  whose two-sided power spectrum $E(\xi)$ satisfyies the Paley-Wiener condition
\begin{equation}\label{PaleyWiener}\int_{-\infty}^{\infty}\frac{|\log E(\xi) |}{1+\xi^2}d\xi<\infty,\end{equation}
can be written as a causal moving average (CMA)
 \begin{equation}Y_h=\langle Y\rangle+\int_{-\infty}^h g(h-s)dW_s,\quad h\in\bbr,\label{CMA}\end{equation}
 where {$W=\{W_{h}\}_{h\in\bbr}$} is a process with uncorrelated and weakly stationary increments, $\langle dW_h\rangle=0$ and $\langle (dW_h)^2\rangle=dh$, with the angle brackets denoting the ensemble average.
 The kernel $g$ is an element of the Hilbert space $L^2$, i.e. $\|g\|_{L^2}^{2}:=\int^{\infty}_0{|g(s)|^2} ds<\infty$ and it is causal, i.e. it vanishes on $(-\infty,0)$.

The converse is also true, i.e. every stationary stochastic process of  form \eqref{CMA} satisfies \eqref{PaleyWiener}.
If we consider $h$ to be time, the representation \eqref{CMA}  has the physically amenable feature of being causal, i.e. $Y$ depends only on the past.
The auto-covariance function of $Y$ has a simple expression: for $\tau\in\bbr$,
 \begin{equation}\label{acf}
\gamma_Y(\tau)=\langle(Y_{h+\tau}-\langle Y\rangle)(Y_h-\langle Y\rangle)\rangle=\int_0^\infty g(s+|\tau|)g(s)ds,
 \end{equation}
while the two-sided power spectrum is \\
$E(\xi)=(2/\pi)\int_0^\infty \gamma_Y(s)\cos(\xi s)ds ={|\mathcal{F}\{g(\cdot)\}|^2(\xi)}$, where
$$\mathcal{F}\{g(\cdot)\}(\xi)=\frac{1}{\sqrt{2\pi}}\int_{0}^\infty g(s)e^{-i\xi s}ds,\quad \xi\in\bbr,$$
 denotes the Fourier operator. 
Condition \eqref{PaleyWiener} is crucial, since for  a general stationary process a representation similar to \eqref{CMA} holds, but the kernel $g$ may not vanish on $(-\infty,0)$ and {the integrals} in \eqref{CMA} and \eqref{acf} are extended {to} the whole real line (\cite{CorrelationTheory}, Ch. 26).
The rate of decay of the spectrum plays a pivotal  role in determining whether a given process has representation \eqref{CMA} or not, since \eqref{PaleyWiener} excludes  spectra decaying at infinity as $\exp(-\xi)$ or faster.
In this context $h$ or $\xi$ do not necessarily denote time or frequency. In the following $x$ will denote a stream-wise spatial coordinate, $\kappa_1$ the associate wavenumber, $t$ a time coordinate, and $\omega=2\pi f$ the associated angular velocity with frequency $f$.

In the turbulence literature the spectral properties of turbulent velocity fields were intensively investigated, starting from Kolmogorov's K41 theory \cite{kolm1941b,kolm1941a,kolm1942}.
In the Wold-Karhunen representation \eqref{CMA} the second-order {properties} of $Y$ depend only on the function $g$, with no necessity to specify the driving noise $W$. {This is analogous to K41 theory, where the second order properties of the velocity field can be handled without considering the intermittent behavior of the turbulent flow.}

From now on, we shall denote the { mean flow velocity component by $V$.
Moreover, we work with the usual Reynolds decomposition $V=U+u$, where $U$ denotes the mean velocity and $u$ is the time-varying part of $V$.}
Then $\re:=UL/\nu$ is the Reynolds number of the flow, by $L$ we denote a typical length, and $\nu$ is the kinematic viscosity of the flow.

In K41  the first universality hypothesis  claims that, for locally isotropic and fully developed turbulence (i.e. $\re\gg1$), the spatial power spectrum $E_L$ of the mean flow velocity fluctuations, has the universal form
\begin{equation}E_L(\kappa_1)=(\varepsilon\nu^5)^{1/4}\Phi_{L}(\eta\kappa_1)=v_\eta^2\eta\Phi_{L}(\eta\kappa_1),\label{universality}\end{equation}
where $\eta=(\nu^3/\varepsilon)^{1/4}$ and  $v_\eta=(\varepsilon\nu)^{1/4}$ are, respectively, Kolmogorov's length and velocity, and $\Phi_{L}(\cdot)$ is {a universal}, a-dimensional function of a-dimensional argument.
{As a cornerstone} of the K41 theory, much effort has been devoted to verify Eq.~\eqref{universality} and to determine the functional form of $\Phi_{L}$. For finite $\re$ we can define {$\Phi_{L}$} in \eqref{universality} as the rescaled spectrum $\Phi^{\re}_{L}$.

In an experimental setting with a probe in a fixed position, the spatial power spectrum can be estimated from the time spectrum $E^\tau_L$, using Taylor's hypothesis \cite{TaylorHP},
\begin{equation}E_L(\kappa_1)= U E^\tau_{L}(U\kappa_1),\label{taylor:hp}\end{equation}
where $\kappa_1=2\pi f/U$.
This is regarded as  a good approximation, when the turbulence intensity $I=\sqrt{\langle u^2\rangle}/U\ll1$ \cite{taylor:error,Lumley}. 
Relation \eqref{taylor:hp} is a first order approximation, where in general higher order corrections are feasible \cite{Gledzer:1997,Lumley}, and also error bounds can be obtained.

Comparison of a large number of experimental low-intensity data sets \cite{local_iso_0,local_iso_1,local_iso_2} shows that in the dissipation range $\Phi^{\re}_L$ is unvarying on a wide range of  Reynolds numbers.
{On the other hand, the inertial range does not exist for small Reynolds number \cite{Champ:Taylor}, however, its length increases with the Reynolds number.}
The supposed infinite differentiability of the solution of the {Navier}-Stokes equation yields $\Phi_L$ to decrease faster than any power in the far dissipation range ($\kappa_1\eta>1$)  \cite{neumann-turb}.
Moreover, the link between second order properties and third order ones is given by Kolmogorov's relation \cite{kolm1941a,pope,Lindborg:1999}
\begin{equation}S_3(x,t)=-\frac{4}{5}\epsilon x+{6\nu}\frac{dS_2}{dx}-\frac{3}{x^4}\int_0^x z^4\frac{dS_2}{dt}dz, \label{kolm:eq}\end{equation}
where $S_n(x,t)=\langle (u(s+x,t)-u(s,t))^n\rangle$ is the $n$-th order longitudinal structure function.
Formula \eqref{kolm:eq} has been used in \cite{turb_spectra_th} to determine an expression for the spatial spectrum which decreases like $\kappa^\alpha_1\exp(-\beta\kappa_1)$ as $\kappa_1$ tends to infinity and $\alpha$ and $\beta$ depend on the Reynolds number.
Numerical   \cite{turb_spectra_sim1,turb_spectra_sim2,turb_spectra_sim3,turb_spectra_sim4,bottleneck:turb} and experimental  \cite{local_iso_1,local_iso_2} studies confirmed such {exponential decay} in the dissipation range.
Under the hypothesis of local isotropy the \emph{local} rate of energy dissipation is
\begin{equation}\epsilon_x={15\nu}(\partial_x V)^2\approx\frac{15\nu}{U^2}(\partial_t V)^2=\varepsilon_t,\label{surrogated2}\end{equation}
where the approximation with the \emph{instantaneous} rate of energy dissipation {$\varepsilon_t$ follows} from Taylor's hypothesis.
As customary in physics literature, we will drop the term rate for the sake of simplicity.
The average energy dissipation can be calculated directly from the spatial spectrum \cite{Batchelor} as
\begin{equation}\epsilon:=\langle\epsilon_x\rangle=15\nu\int_\bbr\kappa_1^2E_L(\kappa_1)d\kappa_1 {\approx \langle\varepsilon_t\rangle},\label{surrogated}
\end{equation}
where the integrand $\kappa_1^2E_L(\kappa_1)$ for $\kappa_1>0$ is the dissipation spectrum.

The generality of the $\CMA$ model \eqref{CMA} requires in the context of turbulence modeling some interpretation of the parameters, in particular, the kernel function $g$. In this paper we estimate the parameters of model \eqref{CMA}, bringing along some physical discussion to motivate our assumptions.

{In Section \ref{section:theory} the model \eqref{CMA}  for the time-wise behavior of the time-varying component $u$ of a turbulent velocity field is motivated first by an analysis of the literature yielding a discussion on the errors of Taylor's hypothesis and consequences for our model.
Physical scaling properties of the kernel lead to a model for $g$ in the inertial and energy containing range.
Moreover, we suggest a deconvolution method to estimate the increments of the driving noise.
In Section \ref{estimation} the estimation methods of Section \ref{section:theory} are applied, using a non-parametric estimation of the kernel on 13 turbulent data sets, having Reynolds numbers $\re$ spanning over 5 orders of magnitude. Moreover the increments of the driving noise for one of the considered datasets is recovered and analyzed.}

\section{Time-wise turbulence model}\label{section:theory}

{In this section we present the theoretical aspects of our turbulence model.
Section 2.1 is devoted to motivating the model by a discussion of Taylor's hypothesis and certain refinements.}
In part Section 2.2 a rescaling of the kernel similar to \eqref{universality} is {suggested}, such that the rescaled kernel depends on the Reynolds number $\re$ and the turbulence intensity $I$ only.
In part Section 2.3 we present a model for the inertial range and the energy containing range in order to deduce the scaling with $\re$ of some features of the rescaled kernel.
Finally, part Section 2.4 deals with the estimation of the increments of the driving noise.

\subsection{Does the Paley-Wiener condition hold?}

First note that an exponentially decaying {power spectrum $E$} violates \eqref{PaleyWiener} and, therefore, it leads to a non-causal representation, regardless of any power-law prefactors.
For spatial spectra this is not against intuition: the basic balance relations leading to the Navier-Stokes equation must hold in every spatial direction; on the other hand, the causality of representation \eqref{CMA} makes sense, when considering the time-wise turbulence behavior.

It is known that the spatial spectra estimated via Taylor's hypothesis \eqref{taylor:hp}  give larger errors in the dissipation range rather than in the inertial range, and that the error is in general positive, i.e. the time spectrum decays at a slower rate than the spatial spectrum. 	{Lumley~\cite{Lumley} derived from a specific model an ordinary differential equation relating time and space spectra. Lumley's ODE was solved in \cite{Champ:Taylor} and used on jet data with $I=0.30$, resulting that the time spectrum obtained via \eqref{taylor:hp} at  $\kappa_1\eta\approx1$ is  238\% higher than the spectrum obtained with Lumley's model}. Moreover, \cite{Gledzer:1997} showed that the {power-law} scaling in the inertial range is substantially left unchanged by Lumley's model. 
A similar effect has been already observed in \cite{Tennekes:1975}, comparing Eulerian and Lagrangian  spectra; such spectra decay as  $\omega^{-2}$ and $\omega^{-5/3}$, respectively.

Based on such facts we postulate that the time spectrum  for every flow with turbulence intensity $I>0$ satisfies \eqref{PaleyWiener}, and it is related to the spatial spectrum by \begin{equation}E_L(\kappa_1)= U E^\tau_{L}(\Lambda_I(U\kappa_1))\label{ex:tai:hp},\end{equation}
where $\Lambda$ depends on $I$, such that $\Lambda_{I}(\kappa_1)/\kappa_1\rightarrow 1$ uniformly in $\kappa_1$ as $I\downarrow 0$.
The classical Taylor hypothesis in \eqref{taylor:hp} assumes that \emph{all} eddies are convected at velocity $U$; however, it is likely that larger eddies propagate with a velocity of the order $U$, whilst the smaller eddies travel at lower velocity, resulting in a less steep decay \cite{moinEditor,alamo}.  The function $\Lambda_I$ accounts for such spectral distortion.
In our framework we will not take the limit $I\downarrow 0$ for two reasons: firstly, since the variance $\langle u^2\rangle$ is finite,  $I\downarrow 0$ would mean that the mean velocity tends to infinity, and, secondly, if  $I\downarrow 0$, in virtue of Taylor's hypothesis, the  time spectrum  \eqref{taylor:hp} would decay exponentially like the spatial one, violating the Paley-Wiener condition \eqref{PaleyWiener}. Therefore, in the considered setting, the limit $I\downarrow 0$ is singular, and we shall consider only the {approximation} for $I\ll1$.
We shall do the same with the singular limit for  $\re$ tending to infinity (\cite{frisch}, Section 5.2), denoting it by $\re\gg1$.

\begin{figure}[htb]
\includegraphics[width=\columnwidth]{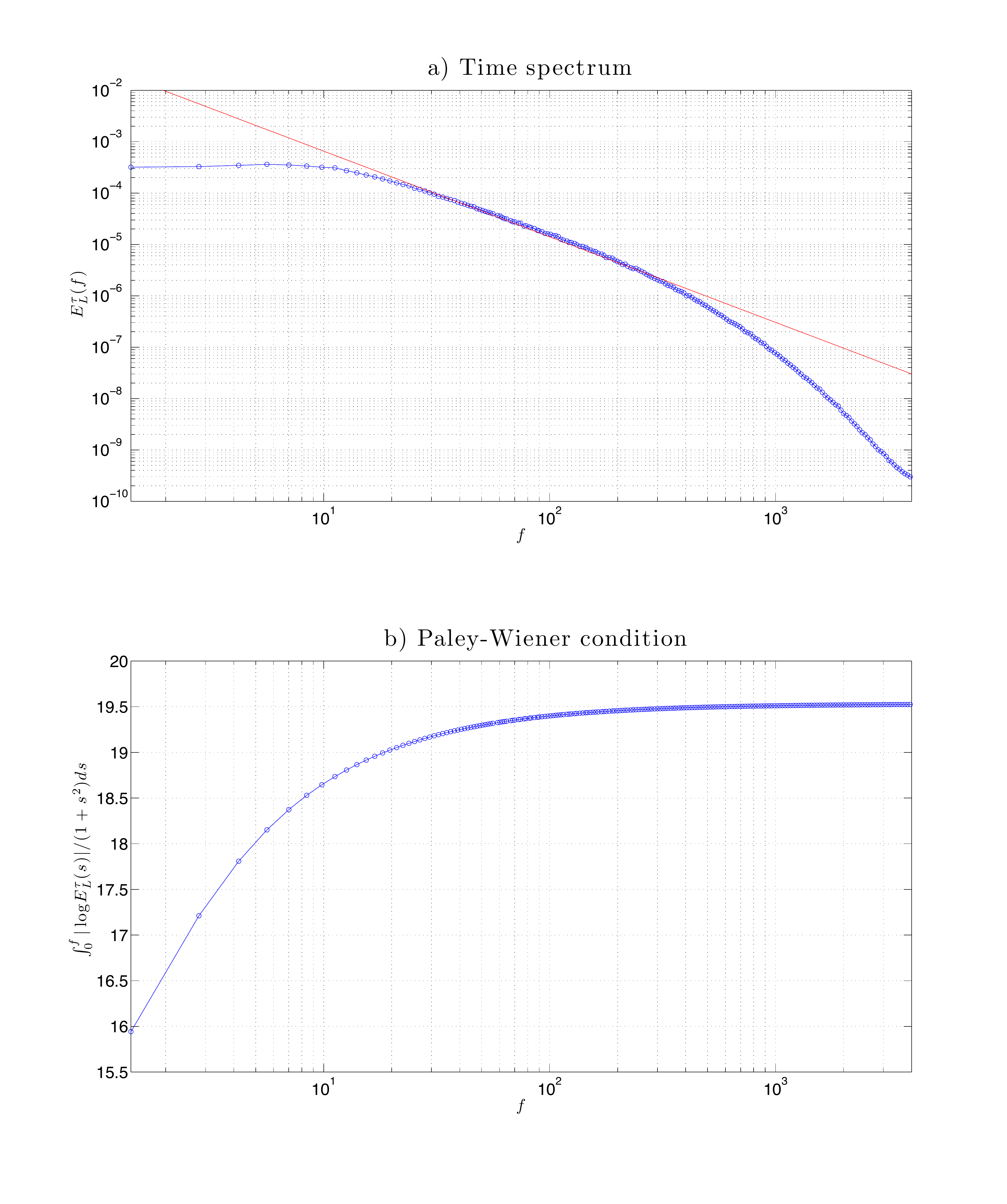}
\caption{a) Time spectrum for data set h3, the solid line indicates Kolmogorov's $5/3$ law. {b) Convergence of the Paley-Wiener integral \eqref{PaleyWiener}. The spectral density is estimated with the Welch method, using a Hamming window of $2^{14}$ data points and $60\%$ overlap.}}
\label{fig:paleyw}
\end{figure}

To date the resolution of experimental data is limited to scales in the order of $\eta$ but, if the data are not too noisy in the dissipation range, it is still possible to check, whether \eqref{PaleyWiener} holds.
In Figure~\ref{fig:paleyw}a) the time spectrum for the data set h3 is depicted; Figure~\ref{fig:paleyw}b) shows that the integral in \eqref{PaleyWiener}  converges and that the dissipation range does not make any significant contribution to the integral.

\subsection{Rescaling the model}

The CMA representation \eqref{CMA} for the time-wise behavior of the mean flow velocity component of a fully developed turbulent flow in the universal equilibrium range (i.e. inertial and dissipation ranges) can be rewritten as
$$V_t=U+C_2\int_{-\infty}^t g_{}(C_1(t-s))dW_s,\quad t\in\bbr,$$
where $g$ is a positive universal function, depending on the Reynolds number $\re$ and the turbulence intensity $I$; $C_1$ and $C_2$ are normalizing constants to be determined.
The integral represents the time-varying part $u$ in the Reynolds decomposition.
The scaling property of the Fourier transform, i.e. that for $c>0$, $c\mathcal{F}\{g(c\,\cdot)\}(\xi)=\mathcal{F}\{g(\cdot)\}(\xi/c)$, gives
$$E^\tau_{L}(\omega)=(C_2/C_1)^2|\mathcal{F}\{g(\cdot)\}|^2\left(\omega/C_1\right).$$
Using \eqref{universality} replacing  $\Phi_L$ with $\Phi^{\re}_{L}$, i.e. without considering the limit for $\re\gg1$, and \eqref{ex:tai:hp}, the relation between the rescaled spatial spectrum and the time spectrum is given by
\begin{equation}v_\eta^2\eta\Phi^{\re}_{L}(\eta\kappa_1)= U(C_2/C_1)^2|\mathcal{F}\{g(\cdot)\}|^2\left(\Lambda_I(U\kappa_1/C_1)\right).\label{taylUniv}\end{equation}

Matching both sides of \eqref{taylUniv}, we get $C_1=U/(2\pi\eta)=f_\eta$, i.e. the Kolomogorov frequency, and $C^2_2={U/v_\eta}\epsilon$.
The Reynolds decomposition can be rewritten as
\begin{equation}V_t=U+\sqrt{\frac{U}{v_\eta}}\int_{-\infty}^t g(f_\eta(t-s))d\tilde{W}_s,\label{rescaled:model}\end{equation}
where $d\tilde{W}_s=\sqrt{\epsilon}{dW_s}$ such that $\langle (d\tilde{W}_s)^2\rangle=\epsilon ds$.
The time-wise increments of the mean flow velocity, at time scale $\Delta$, can be calculated from \eqref{rescaled:model} as
\begin{equation*}
\delta^\Delta_t V=V_{t}-V_{t-\Delta}=\int_{-\infty}^{t-\Delta}[\bar{g}(t-s)-\bar{g}(t-\Delta-s)]d\tilde{W}_s+\int_{t-\Delta}^{t} \bar{g}(t-s)d\tilde{W}_s,
\end{equation*}
where $\bar{g}(\cdot):=\sqrt{U/v_\eta}\ g(f_\eta\cdot)$.
If $\Delta^{-1}\int_{t-\Delta}^{t} \bar{g}(t-s)d\tilde{W}_s\rightarrow 0$ a.s. as $\Delta\downarrow0$,
we have that the derivative process is, with the limit assumed to exist a.s.,
$$\partial_t V=\lim_{\Delta\downarrow 0} \Delta^{-1}\delta^\Delta_t V=\sqrt{\frac{U}{v_\eta}}f_\eta\int_{-\infty}^tg'(f_\eta(t-s))d\tilde{W}_s,$$
i.e. it is again a model of the form \eqref{CMA} {and we essentially exchanged integration and differentiation}.
Moreover, plugging \eqref{universality} into \eqref{surrogated} we obtain
$$\epsilon=15\nu \frac{v^2_\eta}{\eta^2}\int_\bbr s^2\Phi_{L}(s)ds,$$
and, recalling the definition of $v_\eta$ and $\eta$ and \eqref{taylUniv}, we get
\begin{equation}\|\mathcal{F}\{g(\cdot)\}(\Lambda_{I}(\cdot))\,\cdot\|^{-2}_{L^2}=15.\label{L2:time:norm},\end{equation}
The Plancherel theorem, the fact that $$i\xi\mathcal{F}\{g(\cdot)\}(\xi)=\mathcal{F}\{g'(\cdot)\}(\xi)$$ and \eqref{L2:time:norm} give that $\|g'\|^{-2}_{L^2}\approx 15/2\pi$ for $I\ll1$.

{\eqref{surrogated2} gives the instantaneous energy dissipation
\begin{equation}\varepsilon_t:=\frac{15\nu}{U^2}(\partial_t V)^2=\frac{15f_\eta}{2\pi}\int_{-\infty}^t\int_{-\infty}^tg'(f_\eta(t-s_2))g'(f_\eta(t-s_1))d\tilde{W}_{s_1}d\tilde{W}_{s_2}\label{energy:diss},\end{equation}
whose mean value  is  $\varepsilon=\langle\varepsilon_t\rangle=15/2\pi \|g'\|_{L^2}^2 \epsilon\approx \epsilon$ when $I\ll1$.

The constant $U/v_\eta$ is a-dimensional, serving as a scaling factor of the model.
Moreover, the rescaled model \eqref{rescaled:model} indicates that $\{(d\tilde{W}_s)^2\}_{s\in\bbr}$ can account for the observed intermittency, i.e. it must provide all the higher order features of turbulence that can not be reproduced by a Gaussian model, as, for instance, the non-Gaussian behavior of the instantaneous energy dissipation $\{\varepsilon_t\}_{t\in\bbr}$ as indicated by \eqref{energy:diss}. Moreover, we stress that $g$ is a second order parameter of $V$, and so is $g'$.

The model {$d\tilde W_s=\sqrt{\sigma_s} dB_s$} with Brownian motion $\{B_s\}_{s\in\bbr}$ has been suggested in \cite{barndorff-nielsen:372}, where $\sqrt{\sigma_s}$ is the {random} intermittency process,  assumed to be independent of $B$, and $\{\sigma_s\}_{s\in\bbr}$ is the instantaneous energy dissipation, with mean rate $\langle{\sigma_s}\rangle=\epsilon$. 
The major shortcoming of this model is that the Brownian motion assumption implies that the distribution of the increment process  $\{\delta^\Delta_tV\}_{t\in\bbr}$ is symmetric around zero for every scale $\Delta$, which is against experimental and theoretical findings, especially at small scales (see e.g. \cite{frisch}, Section 8.9.3).
Such shortcoming amended in \cite{intermit2} by assuming the presence of a possibly non-stationary drift $Z$, in addiction to $Y$, where $Z$ is smoother than $Y$ (see e.g. \cite{intermit2}, Remark 6). The assumed smoothness of $Z$ implies that the drift has a negligible effect on small scales increments, and is therefore more suitable for modelling phenomena at larger scales, such the energy containing range.

From a second order point of view, everything depends on $U$, $\re$, $\epsilon$ and $\nu$, which are the parameters of \eqref{rescaled:model}. Moreover, given that the dependence of $g$ on $\re$ is known, it is easy to simulate a process having the prescribed time spectrum. As long as only second order properties are of interest, one can indeed take $\{W_s\}_{s\in\bbr}=\{B_s\}_{s\in\bbr}$. If higher order properties are of interest, a realistic model for $W$ is needed.}

\subsection{Dependence of the kernel function  on the Reynolds number}
It has been observed from experimental evidence (see e.g. \cite{frisch}, Section 5.2) that the mean energy dissipation $\epsilon$ is independent of the Reynolds number provided $\re\gg1$.

Since $v_\eta \eta/\nu=1$, as in (4), $U/v_\eta=\re \eta/ L$ holds. From Eq. (6.8) of \cite{pope} we have that $U/v_\eta\propto {\re}^{1/4}$.
From \eqref{acf}, the variance of the mean flow velocity $V$ is
\begin{equation}\label{norm}{\langle u^2\rangle}=\langle(V-U)^2\rangle=v_\eta^2 \|g\|_{L^2}^2,\end{equation}
where  $v_\eta$ is independent of $\re$, when $\re\gg1$, and  $\|g\|_{L^2}$ depends on $\re$ (and to a lesser extent on $I$). Then the turbulence intensity of \eqref{rescaled:model} is, using \eqref{norm},
$$I=\frac{\sqrt{\langle u^2\rangle}}{U}=\frac{v_\eta}{U} \|g\|_{L^2}\propto {\re}^{-1/4} \|g\|_{L^2}.$$
Since $I$ must be independent of $\re$, we have $\|g\|_{L^2}\propto \re^{1/4}$.

A parametric model, suggested in \cite{barndorff-nielsen:372} for the kernel $g$, is the gamma model
\begin{equation}g(t)=C_{\re}  t^{\mu_{\re}-1}e^{- \delta_{\re} t},\quad t\geq0,\label{gamma:model}\end{equation} where $\mu_{\re}>1/2$ and $ \delta_{\re},C_{\re}>0$. This model  yields a power-law time spectrum \cite{pope}
\begin{equation}\label{gamma:spec}E_L^\tau(\omega)=(2\pi)^{-1}{C_{\re}^2\Gamma^2(\mu_{\re})}( \delta_{\re}^2+\omega^2)^{-\mu_{\re}},\end{equation}
where $\Gamma(\cdot)$ is Euler's gamma function.
For instance, the von K\'{a}rm\'{a}n spectrum \cite{karman} is a special case of \eqref{gamma:spec}, where $\mu_{\re}=5/6$ and $C_{\re}=\sqrt{\pi C}/\Gamma(\mu_{\re})\approx1.1431$, and $C$ is the Kolmogorov constant, which has been found to be around $0.53$  over a large number of flows and Reynolds numbers, and therefore universal and independent of the Reynolds number \cite{univKolmConstant}.
For $\omega\ll \delta_{\re}$, the spectrum \eqref{gamma:spec} is constant; hence $\omega\approx \delta_{\re}$ can be interpreted as the \emph{transition frequency} from the inertial range to the energy containing range; moreover, since the upper limit of the inertial range is independent of the Reynolds number, the size of the inertial range varies as $\delta^{-1}_{\re}$.
 
{The gamma model has two essential shortcomings: firstly, it fails to model the steeper spectral decay in the dissipation range, but can be regarded as a good model for the inertial range. Secondly, the kernel has unbounded support, implying that the process is significantly autocorrelated even for large times, although the autocorrelation is  exponentially decaying. 
 
From now on, we shall assume that the kernel function $g$ has compact support, i.e. $g(t)=0$ for $t>T_{\re}$ and $t<0$, where $T_{\re}$ is the decorrelation time \cite{zero:crossing}. Since the inertial range increases with $\re$ (see e.g. \cite{pope}, p. 242) we expect  $\delta_{\re}$ to decrease. For the same reason we expect  $T_{\re}$ to increase with $\re$.}

Explicit computations can be carried out for the truncated gamma model, considering a truncation at $T_{\re}$ and assuming that the failure of the gamma model in the dissipation range does not significantly affect $\|g\|_{L^2}$. Then
\begin{align*}\|g\|^2_{L^2}=&C_{\re}^2\int_0^{T_{\re}} s^{2(\mu_{\re}-1)}e^{-2\delta_{\re} s}ds=\frac{\Gamma \left(2 \mu_{\re}-1\right)-\Gamma \left(2 \mu_{\re}-1,2 T_{\re}\delta_{\re} \right)}{C_{\re}^{-2}2^{2\mu_{\re}-1}} (\delta_{\re})^{1-2\mu_{\re}},\end{align*}
where $\Gamma(a,z):=\int_z^\infty s^{a-1}e^{-s}ds$.
If $T_{\re} \delta_{\re}\gg0$ and it does not vary too much with $\re$, with the choice of the parameters as in the \citeauthor{karman} spectrum and the fact that $\|g\|_{L^2}\propto {\re}^{1/2}$, we get  $\delta_{\re}\propto {\re}^{-3/4}$.

\subsection{Filtering, driving noise and model building}

Let us now consider a second order stationary stochastic process $Y$ with $\langle Y\rangle=0$.
A filtered process $X$ (\cite{priestley}, Eq. 4.12.1) is defined as
$$X_t=\int_0^\infty h(s)Y_{t-s}ds$$ where $h$ is a causal function in $L^1$.
Invoking representation \eqref{CMA} and exchanging the integrals yields
\begin{align}X_t&=\int^\infty_0 h(s)\int_{-\infty}^{t-s}g(t-s-z)dW_{z}ds\notag\\
&=\int_{-\infty}^{t}\int^{t-z}_0 h(s)g(t-z-s) ds dW_{z}\notag\\
&=\int_{-\infty}^{t}(h*g)(t-z) dW_{z}.\label{filtering}\end{align}

We shall assume that $g$ belongs to the so-called Schwartz space $\mathcal{S}$ of rapidly-decaying function; i.e., smooth functions $\phi:\bbr\rightarrow\bbr$ such that $\sup_{s\in\bbr} |s^{p}\partial^{q}\phi(s)|<\infty$ for all integers $p$, $q$. The Fourier transform is well defined for all functions in $\mathcal{S}$, it maps $\mathcal{S}$ onto $\mathcal{S}$ and the inverse Fourier transform of an element of $\mathcal{S}$ is again in $\mathcal{S}$ (\cite{hoermander}, Theorem 7.1.5).
Since in turbulence $|\mathcal{F}\{g(\cdot)\}|$ is assumed to decay faster than any exponential, it belongs to $\mathcal{S}$, and consequently $g\in\mathcal{S}$ as well.

{From \eqref{filtering} we know that the filtering affects the kernel function $g$, but not the driving noise. Moreover, the spectrum of $X$ is, using the convolution theorem,
$$E_X(\omega)=|\mathcal{F}\{(g*h)(\cdot)\}|^2(\omega)={2\pi}|\mathcal{F}\{h(\cdot)\}{\mathcal{F}\{g(\cdot)\}}|^2(\omega),$$
If we know $g$, we can use \eqref{filtering} to build a filter $h$ such that  $X_t\approx\delta^\Delta_t W= W_t-W_{t-\Delta}$.
To this end we choose for appropriate  $\alpha>0$,
\begin{equation}
(h*g)(t)=\phi_\Delta(t)=c_{\alpha,\Delta}e^{-|(t-\Delta) t|^{-\alpha}}\Theta(t)\Theta(\Delta-t),\label{mollifier}\end{equation}
 $c_{\alpha,\Delta}$ is a constant such that $\|\phi_\Delta(\cdot)\|^2_{L^2}=\Delta$, and $\Theta(t)$ is the Heaviside function. 
 Then from \eqref{acf} follows that the filtered process $X$ is a stationary process, with variance $\Delta$ and it is uncorrelated for lag $\tau>\Delta$.
Moreover, $\phi_\Delta(t)$ tends to $\sqrt{\Delta}\delta_{0}(t)$ as $\Delta\downarrow 0$, where $\delta_0(t)$ is the Dirac delta in 0, justifying $X\approx W_{t}-W_{t-\Delta}$ as $\Delta\ll1$. }
 
By the convolution theorem \eqref{mollifier} is equivalent to
\begin{equation}\mathcal{F}\{h(\cdot)\}(\omega){\mathcal{F}\{g(\cdot)\}}(\omega)=\mathcal{F}\{\phi_\Delta(\cdot)\}(\omega),\label{moll:spec}\end{equation}
where saddle-point integration gives that $|\mathcal{F}\{\phi_\Delta(\cdot)\}|(\omega)$ asymptotically decreases like $\omega^{-(\alpha+2)/(2\alpha+2)}\exp(-\omega^{\alpha/(\alpha+1)})$ as $\omega\rightarrow\infty$ and for a fixed $\Delta>0$. We stress that the new kernel $\phi_\Delta(t)$ satisfies the Paley-Wiener condition \eqref{PaleyWiener} for every $\alpha>0$ and, therefore, it is a legitimate kernel function.

W.l.o.g we assume that $\mathcal{F}\{g(\cdot)\}(\omega)\neq0$ for \emph{all} $\omega$, we can deconvolve \eqref{moll:spec}, to get the desired filter $h$. To ensure that $h$ exist, it suffices to choose the parameter $\alpha$ such that
$\mathcal{F}\{\phi_\Delta(\cdot)\}(\omega)/\mathcal{F}\{g(\cdot)\}(\omega)$ is in $\mathcal{S}$ and, as a consequence, that the inverse Fourier transform of such ratio exists. 
{Moreover, {if} the ratio satisfies the Paley-Wiener condition \eqref{PaleyWiener}, then $h$ is causal.}
An ideal choice in the r.h.s. of \eqref{moll:spec} would be  $(h*g)(t)=\Theta(t)\Theta(\Delta-t)$, which can be retrieved for $\alpha\downarrow 0$; however, since $|\mathcal{F}\{g\}|(\omega)$ decays faster than any power of $\omega$, such $h$ does not exist in $\mathcal{S}$.

The procedure shown in this Section is general, and it can be applied without restriction to any physically meaningful process, where a model of the form \eqref{CMA} applies. 

\section{Estimation}\label{estimation}

\subsection{The data}
\begin{table*}
{\hskip -1.2cm  \scriptsize
 \begin{tabular}{|c|ccccccc|ccccc|}\hline
Data& $\epsilon\cdot 10^2$ &$\eta$&$\lambda$&$\rl$&$F$&$f_\eta$&$I$&$\|g\|_{L^2}$&$\mu_{\re}$&$ \delta_{\re}\cdot10^{4}$&$T_{\re}\cdot 10^{-2}$&$C_2$\\
&$m^2s^{-3}$&$\mu m$&$mm$&&$kHz$&$kHz$&$\%$&&&&&\\
\hline\hline
h1 &2.72&36.42&0.76&112&5.74&1.14&20.56&5.37&0.76&363.85&0.98&0.84\\
\hline
h2 &0.52&55.07&1.11&105&2.87&0.52&19.19&5.21&0.73&387.11&1.05&0.38\\
\hline
h3 &20.14&22.08&0.55&162&22.97&3.57&21.47&6.46&0.85&276.51&1.38&2.46\\
\hline
h4 &0.85&15.7&0.49&253&5.74&1.74&24.04&8.08&0.81&120.81&3.06&0.53\\
\hline
h5 &13.14&24.56&0.66&184&45.94&6.01&10.98&6.9&0.95&316.1&1.25&2.87\\
\hline
h6 &10.98&8.28&0.36&495&22.97&9.02&23.34&11.31&0.83&63.57&5.95&2.31\\
\hline
h7 &64.37&5.32&0.26&640&91.88&25.66&22.58&12.85&0.87&49.87&7.53&6.05\\
\hline
h8 &1995.62&2.25&0.14&930&367.5&175.68&22.15&15.5&0.84&22.95&13.81&37.37\\
\hline
h9 &2120.81&2.22&0.14&978&367.5&190.05&21.64&15.89&0.84&21.78&15.79&39.46\\
\hline
h10 &5192.01&1.78&0.11&1005&367.5&285.13&22.88&16.11&0.83&18.13&18.42&60.46\\
\hline
h11 &13580.72&1.4&0.1&1336&735&570.02&21.34&18.58&0.8&11.76&26.57&108.72\\
\hline
a1 &1.38&837.47&140.01&7216&10&1.21&15.29&43.16&0.77&0.55&282.72&1.97\\
\hline
a2 &9.41&442.42&115.85&17706&5&2.99&28.23&67.61&0.81&0.24&882.39&4.75\\
\hline
 \end{tabular}}
 \caption{ h1-h11: Helium data \cite{heliumData}, a1: Sils-Maria data \cite{SilsMaria},  a2: Brookhaven data \cite{Dhruva}. The norm $\|g\|_{L^2}$ is evaluated using \eqref{norm}.}
 \label{tab1}
\end{table*}
In this section we estimate the kernel function {$g$ non-parametrically}, the parameters of model \eqref{rescaled:model}, and the increments of the driving noise $W$.
Not surprisingly, the quality of the data in the  present study does not allow us to perform a reliable estimation in the dissipation range. For the kernel in the universal equilibrium range, the gamma model \eqref{gamma:model} is estimated by the non-parametric kernel estimation method as suggested in \cite{bfk:2011:2}.
 We analyzed 13 different data sets, whose characteristics are summarized in Table \ref{tab1}. Also dependence of the parameters on $\re$ is considered.
Two of the data sets examined here come from the
atmospheric boundary layer (a1 \cite{SilsMaria} and a2 \cite{Dhruva})
and eleven from a gaseous helium jet flow (records h1 to h11  \cite{heliumData}).
Since the data sets come from different experimental designs, characteristic quantities
such as Taylor's microscale Reynolds number $\rl=\sqrt{\langle u^2\rangle}\lambda/\nu$ based on
Taylor's microscale $\lambda=\sqrt{15\langle u^2\rangle\nu/\epsilon}$ are considered.
This Reynolds number is unambiguously defined and  the  rough estimate $\rl\propto\sqrt{2\re}$ holds (\cite{pope}, p. 245). In all data sets  the turbulence intensity $I$ is rather high and, therefore, the expectation of the r.h.s of \eqref{surrogated2} would not be a good approximation for $\epsilon$.
Moreover, in most of the data sets the inertial range is hard to identify due to low Reynolds numbers.
The mean energy dissipation $\epsilon$ has been estimated from \eqref{kolm:eq}, following \cite{Lindborg:1999}.
Estimation of $\epsilon$ is a challenging and a central task, since the rescaling constants in \eqref{rescaled:model} depend, in the K41 spirit, only on $\epsilon$, $\nu$ and $U$.
\begin{figure}[h]
\hskip -0.64 cm\includegraphics[width=1.15\columnwidth]{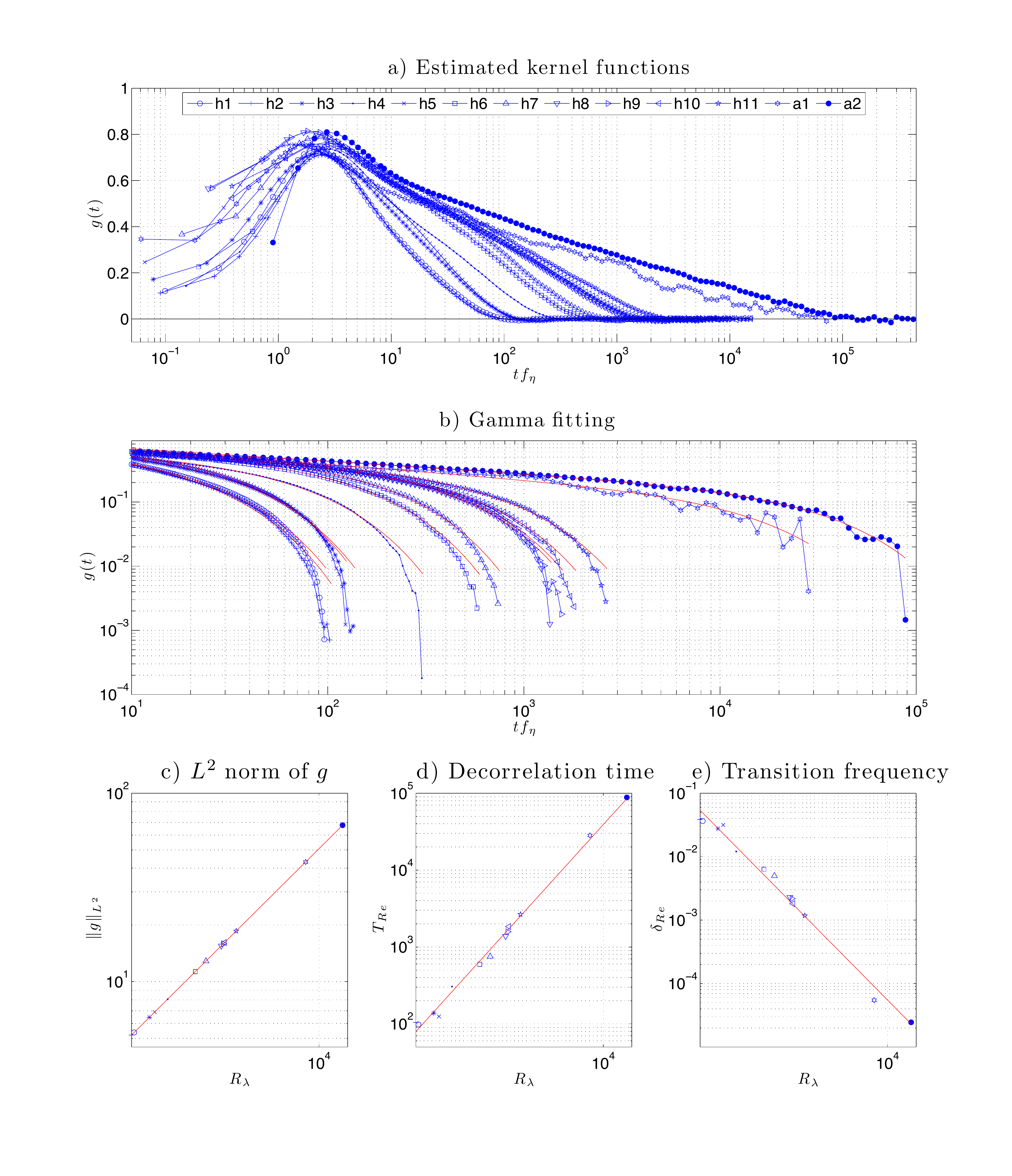}
\caption{a) Estimated kernels, against logarithmic time scale. b) Estimated kernels, plotted in log-log scale. Solid lines are the fitted gamma models \eqref{gamma:model}. c-e) Decorrelation times $T_{\re}$, transition frequency $\delta_{\re}$ and $\|g\|^2_{L^2}$  as a function of Taylor's microscale-based Reynolds number. Solid lines represents  power-law fittings of the considered quantities vs. $\rl$.}
\label{fig:kernels}
\end{figure}

\subsection{Kernel estimation}\label{kern:est}

The kernel function $g$ {characterizes} the second order properties of the mean flow velocity $V$, i.e. spectrum, autocovariance and second order structure function. Classical time series methods have been employed to estimate the kernel function of high frequency data \cite{bfk:2011:2} with a non-parametric method. The method is essentially based on finding the unique function $g$ solving \eqref{acf}, when the autocovariance function $\gamma_Y$ is given. {We estimate the discrete autocovariances} $\gamma^\Delta_Y(i):=\gamma_Y(i\Delta)$, where $\Delta$ is the sampling grid size and $i$ are integer values.
It has been shown {\cite{bfk:2011:2}} that the coefficients of a high order discrete-time moving average process fitted to such closely observed autocovariance, when properly rescaled, estimate consistently the kernel function on the mid-point grid; i.e, $\hat{g}(i)\approx g((i+1/2)\Delta)$ for integers $i$.

 The results are presented in Figure \ref{fig:kernels}a), where all the estimated kernels are plotted on a logarithmic time axis. Due to instrumental noise, the high turbulence intensity of {the data sets} and the non-infinitesimal dimension of the hot-wires, estimates in the dissipation range ($tf_\eta<1$) {differ} significantly from one data set to another. In general {we notice} that the steeper decay of the spectrum in the dissipation range is reflected by the fact that the kernel functions {tend} to 0 for $tf_\eta\ll1$, instead of exploding as the gamma model for $\mu_{\re}=5/6$. The only estimation of the kernel that is not perfectly aligned with the others is the one for h5, which looks slightly shifted to the left. This can be attributed to the difficulty in estimating $\epsilon$, and consequently $f_\eta$.
In Figure~\ref{fig:kernels}c) the estimated $\|g\|_{L^2}$ based on \eqref{norm} are plotted against $\rl$, and the power-law fitting shown in the figure returned $\|g\|_{L^2}=0.5081\rl^{0.5}\propto{\re}^{0.25}$.

The {decorrelation} time {$T_{Re}$} has been estimated by the first zero-crossing of the estimated $g$.
It increases with $\rl$ (Figure  \ref{fig:kernels}a)) and it follows empirically the law $T_{\re}= 0.1408\rl^{1.3613}\propto\re^{0.6806}$  (Figure  \ref{fig:kernels}c)). 

The transition frequency $\delta_{\re}$ is obtained via {least-squares} fitting {of} the gamma model \eqref{gamma:model} to the non-parametric estimate of $g$ for $t f_\eta>5$ (Figure~\ref{fig:kernels}b)). The statistical fits in Figure~\ref{fig:kernels}b) are very good for all data sets considered, at least when not too close to  $T_{\re}$.
The transition frequency follows the empirical law $ \delta_{\re}=61.4917\rl^{-1.5127}\propto\re^{-0.7564}$ (Figure~\ref{fig:kernels}e)), which {is} exceptionally close to the exponent $-3/4$ {estimated} {for} the gamma model, and $\delta_{\re} T_{\re}$ is between 1.5 and 4.1 in all data sets.
The estimated values of $\mu_{\re}$ {are} also close to the reference value of $5/6\approx 0.834$, with a mean value of 0.8218. The only notable outlier is the data set h5, which has been already proven to be somewhat anomalous.

As said in Section 2.3, the truncated gamma model is not able to capture the sharp cutoff in the neighborhood of $T_{\re}$ nor the rapid decrease in the dissipation range.
%
{To estimate how the variance of the model $\eqref{CMA}$ is affect by the truncation at $T_{\re}$, we consider the quantity
\begin{equation*}H(\re)=\frac{\| g(\cdot)\Theta(T_{\re}-\cdot) \|^2_{L^2}}{\| g(\cdot) \|^2_{L^2}}=1-\frac{\Gamma (2/3,2 c_1 c_2 {\re}^{\beta -\alpha })}{\Gamma \left(2/3\right)}\end{equation*}
where the latter is obtained  using the gamma kernel \eqref{gamma:model} with parameters $\mu_{\re}=5/6$, $T_{\re}=c_1{\re}^{\beta}$ and $\delta_{\re}=c_2\re^{-\alpha}$, where $c_1,c_2>0$ and $\alpha>\beta>0$, as indicated by the least-squares in Figure \ref{fig:kernels}. $H$ represents the ratio between the variance \eqref{norm} using a truncated gamma model and a non-truncated one, and it is a decreasing function of $\re$, and it tends to 0 as $\re\gg1$, indicating that the truncation is important when  $\re\gg1$; i.e., when $T_{\re}\delta_{\re}\ll1$.
Nonetheless, using the values of $c_1,c_2,\alpha,\beta$ returned by the least-squares fitting, we obtain $H\approx 0.991$ for the dataset a2, which exhibits the highest Reynolds number among the considered datasets.
Then we can conclude that, for the wide range of Reynolds numbers considered, the variance of the model \eqref{CMA} with a kernel following  the truncated gamma model does not differ in a sensible way from the variance of the classical von K\'{a}rman model. 
 
Finally, the behavior of $\delta_{\re}$ as function of $\re$ estimated via the gamma model agrees significantly with the data, showing that the dissipation range does not contribute appreciably to $\|g\|_{L^2}$, which is in agreement with the idea that there is few energy in the dissipation range.}

\subsection{Noise extraction}

Relation \eqref{moll:spec} was calculated for a  continuous time process with the goal of {estimating} discrete time increments of the driving noise.
In practice, if we have data sampled with grid size $\Delta$, we can reconstruct the sampled quantities in the frequency domain only on  $0\leq f<F/2=(2\Delta)^{-1}$.
If $\Delta\ll1$, we can use the approximation $\mathcal{F}\{\phi_\Delta(\cdot)\}(\omega)\approx1$ for $\omega\leq \pi/\Delta$;
similarly, for the Fourier transforms the relation $\mathcal{F}\{h(\cdot)\}(2\pi k/(N\Delta))= \Delta DFT\{h^\Delta\}(k)+o(\Delta)$ holds (\cite{numericalrecipes}, Eq. 13.9.6), for $N/2+1<k<N/2$, $h^\Delta=\{h(i\Delta)\}_{i=1,\ldots,N}$, i.e., it is a function $h$ sampled with {grid size} $\Delta$ and $DFT$ is the discrete Fourier transform. The number of {observations} of the mean flow velocity $V^\Delta$ is denoted by $N$.
Then the increments of the driving noise on a discrete grid can be recovered {by} computing
\begin{equation}\label{discrete:decon}\delta^\Delta_{i\Delta}W\approx IFT\left\{\frac{DFT\{V^\Delta\}(k)}{{DFT\{\hat{g}\}(k)}}\right\}(i),\end{equation}
where $\hat{g}$ is the {non-parametrically} estimated $g$, truncated at $T_{\re}$ and properly zero-padded to increase its length to $N$. Moreover, $IFT$  denotes the inverse discrete Fourier transform. For turbulence, the high-frequency condition $\Delta\ll1$ can be regarded as satisfied whenever $(2\Delta)^{-1}=F/2< f_\eta$; i.e., if the resolution of the data is of the order of the Kolmogorov's length.
\begin{figure}[h!]
\hskip -1.2 cm \includegraphics[width=1.3\columnwidth]{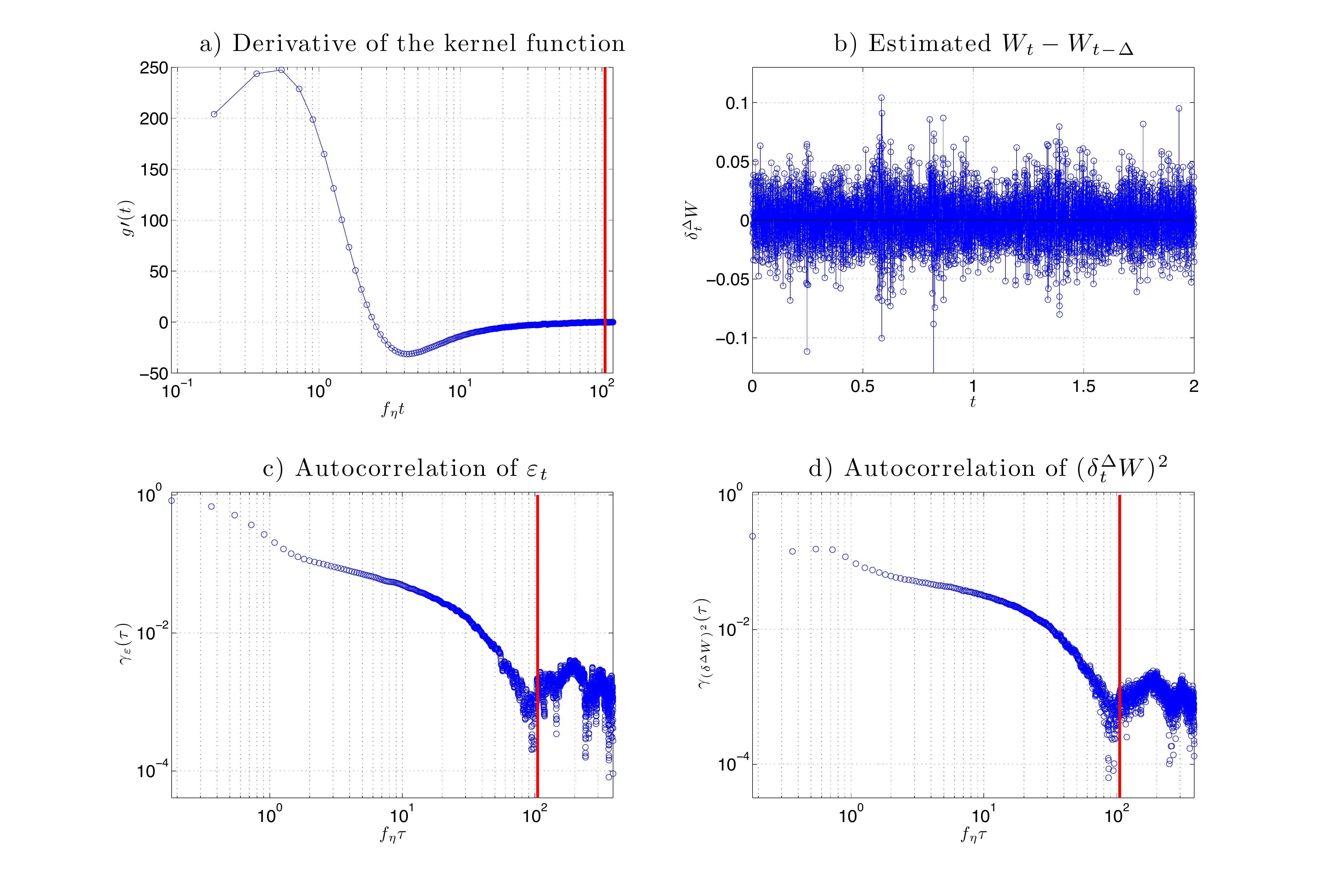}
\caption{a) Derivative of the kernel. b) Increments of the driving noise, estimated with formula \eqref{discrete:decon}. c) Autocorrelation function of the instantaneous energy dissipation $\varepsilon_t$. d) Autocorrelation function of the squared noise $(\delta^\Delta_{i\Delta}W)^2$. The vertical solid lines in a), b), d) denote the decorellation time $t=T_{\re}$.}
\label{fig:noise}
\end{figure}
\begin{figure*}[h!]
\includegraphics[width=1.1\columnwidth]{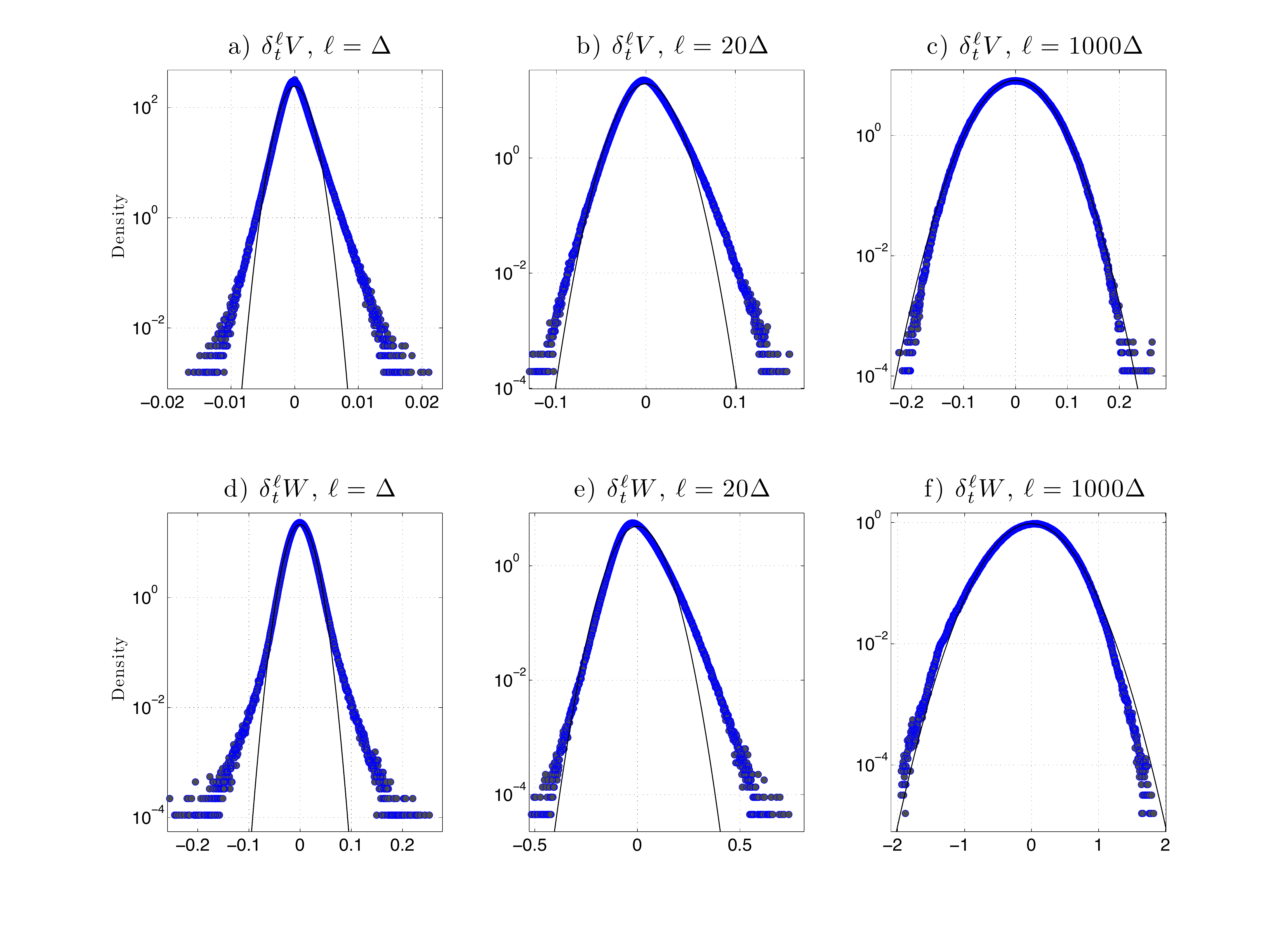}
\caption{a-c). Histogram of the increments of the main-flow velocity for the dataset h2, respectively at scales $\ell=\Delta,20\Delta,1000\Delta$.
d-f) Histogram of the estimated increments of the intermittency process for the dataset h2, respectively at scales $\ell=\Delta,20\Delta,1000\Delta$.}
\label{fig:scaling}
\end{figure*}
{We apply \eqref{discrete:decon} to the dataset h2, which shows low noise level and good resolution in the dissipation range  ($F/f_\eta\approx5.52$); in Figure~\ref{fig:noise}b) part of the obtained increments are plotted as an example.  The increments of the driving noise show a clear intermittent behavior, exhibiting clustering and a non-Gaussian distribution. The clustering, i.e. the fact that larger increments are not isolated but appear in clusters, is reflected in the autocorrelation function of  $(\delta_t^\Delta W)^2$, showed in Figure~\ref{fig:noise}d), which is very similar to the one of $\varepsilon_t$ (Figure~\ref{fig:noise}c)), defined in \eqref{surrogated2}. It is also remarkable that those two autocorrelations are in the order of $10^{-3}$ for $tf_\eta>T_{\re}=105$, suggesting that that $V_{t}$ may be independent of $V_{t-\tau}$ for $\tau>T_{\re}$, rather than simply uncorrelated. Similar plots can be obtained for the other datasets as well.

The similarity between those two autocorrelation function can be explained by the fact that the derivative of the kernel function, plotted in Figure~\ref{fig:noise}a), is concentrated in $0<tf_{\eta}<2$, it tends rapidly to zero for $t\downarrow 0$ and is small for $tf_\eta>30$. Then we can heuristically think of approximate $g'(t)$ with $\Theta(t-\tilde{T}_1)\Theta(\tilde{T}_2-t)$ for some $0<\tilde{T}_1<\tilde{T}_2\ll T_{\re}$, then  \eqref{energy:diss} reduces to 
$$\varepsilon_t\approx \frac{15f_\eta}{2\pi}(W_{t-\tilde{T}_1}-W_{t-\tilde{T}_2})^2.$$ 
Moreover the increments of the intermittency process shows a density similar to the increments of the velocity, with positive skewness and exponential tails (Figure~\ref{fig:scaling}a) and d)) at small scales, and looking more and more Gaussian as the scales increase (Figure~\ref{fig:scaling}b-c) and Figure~\ref{fig:scaling}e-f)). These phenomena are universally recognized as distinguished features of turbulence (\cite{frisch}, Ch. 8).}


\section{Discussion and conclusion}
In this paper we proposed application of a large class of stochastic processes in the context of time-wise turbulence modelling. The class of CMA processes \eqref{CMA} is rather flexible, with the only constraint of having a spectral density satisfying the Paley-Wiener condition \eqref{PaleyWiener}, which excludes processes with spectrum decaying too fast but it still allows them  to have sample path infinitely differentiable with probability 1. Although causality is a reasonable feature of time-wise behavior of turbulence, it is not in the spacial one, and the link between the two is the often criticized Taylor's frozen field hypothesis.  To the authors' knowledge, this is the first time the question is raised and it is worthy to be studied in more detail, e.g. using DNS simulation as in \cite{alamo}.

Essentially the CMA model distinguishes between second order properties, accounted by the kernel function, and higher order ones, depending on the noise, which can be specified independently from each other, in agreement with the K41 theory. The dependence of the kernel function on the Reynolds number is analyzed, with a special regard to its behavior far away from the origin. The analysis in the time domain allows higher resolution of second order properties at higher lags, which corresponds to the leftmost part of the one-sided spectrum, showing in a clear way how the inertial range, proportional to $\delta^{-1}_{\re}$, and the decorellation time $T_{\re}$ increase with the Reynolds number. 

We propose a modification of the gamma model of \cite{barndorff-nielsen:372}, with the parameters depending explicitly on the Reynolds number, modelling well the inertial range and the transition to the energy range.
Unfortunately the present data, due instrumental noise, does not allow a precise analysis of the behavior of the kernel function near to the origin, which corresponds to the dissipation range. Such an analysis would be possible using data coming from computer simulations, and it is left for future research.

Moreover, a method to recover the driving noise is proposed, essentially based on constructing a linear operator to suppress the linear dependence of the data. The obtained noise, which is dimensionally the square root of the energy dissipation, shows some of the features of the energy dissipation $\varepsilon_t$, collectively known as intermittency.  That shows that the second order dependence does not play an important role in determining the high order statistics.

We conclude mentioning that the analysis performed in this paper holds in great generality for one-dimensional processes. It is possible to model the full three-dimensional, time-wise behavior of turbulent velocity field with a similar CMA model, and under the hypothesis of isotropy a model for the three-dimensional kernel function can be obtained from a one-dimensional one, since the two point correlator depends only on the longitudinal autocorrelation function (see e.g. \cite{pope}, p. 196).
\section{Acknowledgment}

 {We are grateful to J\"urgen Schmiegel (Aarhus University), Benoit Chabaud (Joseph Fourier University, Grenoble) and Beat L\"uthi (ETH Zurich) for sharing their data with us.} V.F.'s work was supported by the International Graduate School of Science and Engineering (IGSSE) of the Technische Universit\"at M\"unchen. C.K. gratefully acknowledges financial support by the TUM Institute for Advanced Study (TUM-IAS).

\bibliographystyle{model5-names}
\bibliography{bib}







\end{document}